\documentclass[prl, twocolumn, amsmath, amssymb, notitlepage, longbibliography]{revtex4-1}

\usepackage{dsfont}
\usepackage[english]{babel}
\usepackage{graphicx}
\usepackage{amsmath}
\usepackage{hyperref}
\usepackage[usenames, dvipsnames]{color}

\graphicspath{ {./Pictures/} }

\newcommand{\bra}[1]{\ensuremath{\left\langle #1\r|}}
\newcommand{\ket}[1]{\ensuremath{\left|#1\r\rangle}}

\newcommand{\mean}[1]{\ensuremath{\left\langle #1\r\rangle}}

\newcommand{\hc}{^{\dagger}}             					% suffix for the hermitian conjugate
\newcommand{\HC}{\textrm{h.c.}}
\newcommand{\ee}{e}             					% exponential function
\renewcommand{\L}[0]{\mathcal{L}}  						% Lagrangian

\newcommand{\comm}[2]{\left[ #1, #2 \right]} 				% commutator
\newcommand{\nn}{\nonumber}							% nonumber abreviation
\newcommand{\abss}[1]{\ensuremath{ \left| #1 \right|^{2} }}	% absolute value squared
\newcommand{\diss}[1]{\mathcal{D}[ #1 ]}					% dissipator
\renewcommand{\l}[0]{\left}
\renewcommand{\r}[0]{\right}

\newcommand{\Tr}{\text{Tr}}

\newcommand{\eqn}[1]{Eq.~\ref{#1}}
\newcommand{\fig}[1]{Fig.~\ref{#1}}

\begin{document}

\title{Dephasing-assisted Gain and Loss in Mesoscopic Quantum Systems}

\author{Clemens M\"uller}
\affiliation{ARC Centre of Excellence for Engineered Quantum Systems, School of Mathematics and Physics, University of Queensland, Saint Lucia, Queensland 4072, Australia}

\author{Thomas M. Stace}
\affiliation{ARC Centre of Excellence for Engineered Quantum Systems, School of Mathematics and Physics, University of Queensland, Saint Lucia, Queensland 4072, Australia}

\date{\today}

\begin{abstract}
Motivated by recent experiments, we analyse the phonon-assisted steady-state gain of a microwave field driving a  double quantum-dot in a resonator.  We apply the results of our companion paper, which derives the complete set of fourth-order Lindblad dissipators using  Keldysh methods, to show that resonator gain and loss are substantially affected by dephasing-assisted dissipative processes in the quantum-dot system.  These additional processes, which go beyond recently proposed polaronic theories, are in good quantitative agreement with  experimental observations.
\end{abstract}

%\keywords{}
%\pacs{}

\maketitle

%%%%%%%%%%%%
% Intro
%%%%%%%%%%%%

Microwave-driven double-quantum dots (DQD) have demonstrated a rich variety of quantum phenomena, including population inversion~\cite{pet04,Stace:PRL:2005, Petta:S:2005,Stace:PRL:2013, Colless:NC:2014}, gain~\cite{Stockklauser:PRL:2015, Liu:PRL:2014, Kulkarni:PRB:2014}, 
masing~\cite{Marthaler:PRL:2011, Jin:RPP:2012, Liu:S:2015, Liu:PRA:2015, Marthaler:PRB:2015, Karlewski:PRB:2016} and Sysiphus thermalization~\cite{Gullans:PRL:2016}. 
These processes are well understood in quantum optical systems, however mesoscopic electrostatically-defined quantum dots exhibits additional complexity not typically seen in their optical counterparts, arising from coupling to the phonon environment.  

\begin{figure}
	\begin{center}
	\includegraphics[width=\columnwidth]{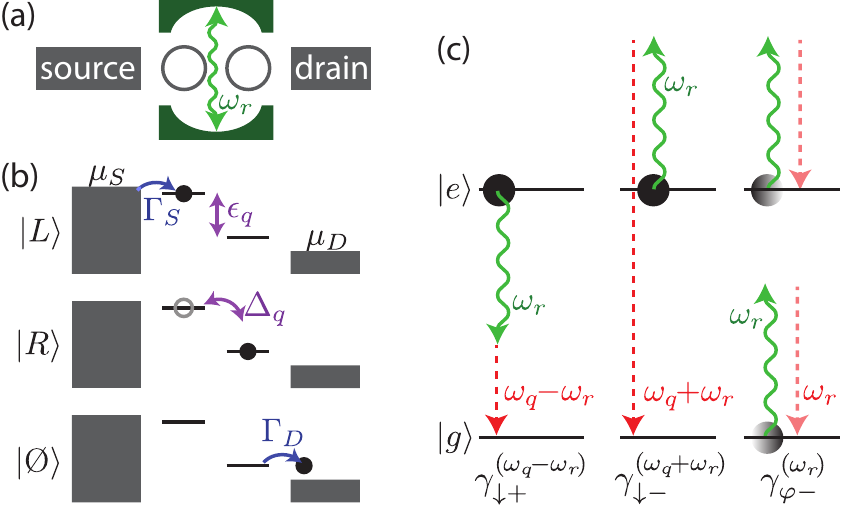}
	\caption{(a) Schematic showing  a double quantum dot (DQD) coupled to a cavity resonator mode, and to source/drain leads. 
		(b) An electron  (black dot) tunnels from the source lead to the DQD state $\ket{L}$, which couples to the DQD state $\ket{R}$ with matrix element $\Delta_q$, and then tunnels to the drain, leaving the DQD in the empty state $\ket{\O}$. 
 		Also shown are inter-dot bias, $\epsilon_q$, and coupling rates $\Gamma_{S,D}$ to metallic leads, with chemical potentials $\mu_{S,D}$.   (c) Dissipative processes due to phonon emission (dashed  arrows) responsible for rates in~\eqn{eq:CorrelatedRates}.  
		Each processes is correlated with resonator photon  creation (downward  wiggly arrows) or annihilation (upward  wiggly arrows); $\gamma_{\downarrow\pm}^{(\omega_q\pm\omega_r)}$ correspond to DQD relaxation from $\ket{e}$ to $\ket{g}$, 
		whilst  $\gamma_{\varphi-}^{(\omega_r)}$ corresponds to DQD dephasing, leaving the  populations of $\ket{e}$ and $\ket{g}$ unchanged.
	} 
	\label{fig:dissipation}
	\end{center}
\end{figure}

A notable experimental example of this, which motivates our work, is an  electronically open, DQD system  coupled to a driven resonator, pictured in~\fig{fig:dissipation}a,b,~\cite{Liu:PRL:2014}.  
Substantial gain in the resonator field  was observed when the DQD is blue-detuned with respect to the resonator, and capacitively biased to induce substantial population inversion.  
The observed gain is attributed to correlated emission of a resonator photon and a phonon into the semiconductor medium in which the DQD system is defined.  
This process ensures conservation of energy, since the phonon carries the  energy difference, $\hbar(\omega_q-\omega_r)$, between the energy of the qubit  and the energy of the resonator phonon (we set $\hbar=1$ for the rest of the letter).  

In some experimental regimes of~\cite{Liu:PRL:2014}, the observed gain is well described by a theory  based on a canonical transformation to a polaron frame~\cite{Gullans:PRL:2015}.  
In this frame, conventional quantum optics techniques and approximations   (Born-Markov, secular etc) are used to derive dissipative Lindblad superoperators that are quadratic in both the qubit-phonon bath coupling strength, $\beta_j$, and  in the qubit-resonator coupling strength, $g$.    
However the same theory fails to describe substantial loss (sub-unity gain) in other experimental regimes, which strongly suggests that there are additional dissipative processes that are not captured in the polaron frame.

One problem with relying on a canonical transformation as the basis for a perturbative expansion is that it is  tailored to a specific  frame, which emphasises some processes over others. 
It is therefore not guaranteed to find all dissipative processes that  occur at a given order in perturbation theory.

In a longer technical companion paper \cite{Mueller:PRA:2016}, we consider a generic two-level system (qubit) coupled to a resonator, and to a bosonic bath, and we present a  derivation of the complete set of Lindblad superoperators that arise at the same order as those in~\cite{Gullans:PRL:2015}, 
i.e.\ $\sim O(\beta_j^2 g^2)$.  Our derivation is based on Keldysh diagrammatic perturbation theory, and makes explicit the nature of the approximations and idealisations we deploy.  
In Ref.~\onlinecite{Mueller:PRA:2016} we show that as well as the correlated decay processes in Ref.~\onlinecite{Gullans:PRL:2015}, we find a number of additional Lindblad dissipators at the same perturbative order.  
Amongst these additional dissipators, there is a process describing correlated dephasing of the qubit accompanied by resonator photon emission or absorption.  
This leads to additional qubit-mediated resonator gain and loss terms, which we believe have not been derived previously in the Lindblad formalism.

In this Letter, we apply the results of our companion paper~\cite{Mueller:PRA:2016} to the specific problem of the open DQD system  coupled to a driven resonator and to a phonon environment.  
We show that the additional Lindblad dissipators that arise in the Keldysh analysis of the Dyson series do indeed generate substantial additional loss, and that the resulting theory consistently accounts for the  magnitude of the gain and loss in different experimental regimes.

%%%%%%%%%%%%
% System
%%%%%%%%%%%%
%\paragraph{System} - 
In what follows, the `system' consists of the resonator and the DQD (pictured in~\fig{fig:dissipation}a), which is driven at frequency $\omega_{d}$ and amplitude $\epsilon_{d}$,  and the `bath' is the  environment of phonon modes, $b_j^\dagger$.  Interactions between the DQD and the resonator photons and bath phonons will be treated perturbatively.   
As such, we partition the  total Hamiltonian as $H=H_S+H_D+H_B+H_I$, where
\begin{eqnarray}
	H_{S} &=& \omega_{r} a\hc a -  \epsilon_{q} \sigma_{z}^{(p)}/2 +  \Delta_{q} \sigma_{x}^{(p)}/2, \nn\\
	H_D &=&  \epsilon_{d} \l( a\hc \ee^{ i \omega_{d} t} + a \ee^{- i \omega_{d} t} \r)/2,\nn\\
	H_B&=&{\sum}_{\textrm{modes }j}\omega_{j} b_{j}\hc b_{j},\nn\\
	H_I&=&g \sigma_{z}^{(p)} ( a + a\hc )/2+ \sigma_{z}^{(p)} X /2,\nn
\end{eqnarray}
where $\sigma_{z}^{(p)} ={ \ket R}{ \bra R} - {\ket L}{ \bra L}$ and \mbox{$\sigma_x^{(p)}={\ket R }{\bra L} + {\ket L}{ \bra R}$} are DQD operators expressed in the position basis $\{{\ket{L}},{\ket{R}}\}$ (shown in~\fig{fig:dissipation}b), %\mbox{$\sigma_x^{(p)}={\ket R }{\bra L} + {\ket L}{ \bra R}$}, %induces transitions between the two dots 
 $a$ is the resonator annihilation operator, and \mbox{$X=\sum_{j}\beta_{j} ( b_{j} + b_{j}\hc )$} is the phonon coupling operator.

We transform to an interaction frame defined by \mbox{$H_0=H_S+H_B$}, so that the interaction Hamiltonian in the DQD energy eigenbasis becomes
\begin{align}
	H_{ I}(t) =&   g \left( \cos\theta\: \sigma_{z} + \sin\theta\: \sigma_{x}(t)  \right) \left( a e^{-i\omega_r t}+ a\hc e^{i\omega_r t} \right)/2 \nn\\
		& \hspace{-8mm}+  \left( \cos\theta\: \sigma_{z} + \sin\theta\: \sigma_{x}(t) \right) X(t)/2,\label{eqn:HI}
\end{align}
where {$\sigma_z={\ket g}{ \bra g} -{\ket e}{\bra e}$}, {$\sigma_x(t)=e^{i\omega_q t}{\ket e }{\bra g} + \HC$}, $\tan{\theta} = \Delta_{q} / \epsilon_{q}$,  \mbox{$\omega_{q} = ({\epsilon_{q}^{2} + \Delta_{q}^{2}})^{-1/2}$} is the DQD energy splitting, and \mbox{$X(t)=\sum_{j}\beta_{j} ( b_{j} e^{-i\omega_j t} + b_{j}\hc e^{i\omega_j t} )$}.

We have made a rotating wave approximation in $H_D$, and  we also assume weak coupling between the system and bath. Further, in the experiment we consider, $g\ll |\omega_{q} -\omega_{r}|$.   
Within these approximations we derive a Lindblad master equation in the interaction Hamiltonian $H_I(t)$, using  Keldysh perturbative methods.  Different dissipative processes arise at different orders of the DQD-resonator and DQD-phonon coupling strengths.

The dynamics of the open DQD-resonator system are governed by the master equation 	
\begin{align}
	\dot \rho &=-i[H_D(t),\rho] + \L_{2} \rho+\L_{4} \rho + \L_{\text{leads}}\rho. 
	 \label{eq:QDBias}
\end{align}
In this expression, the resonator driving hamiltonian coherently populates the resonator, and the DQD-phonon and DQD-resonator coupling gives rise to dynamical superoperators $ \L_{2}$ and $ \L_{4}$ at different orders of perturbation theory in the coupling strengths $g$ and $\beta_j$. 
Electronic coupling to the leads gives rise to the dissipators $\L_{\text{leads}}$.  In what follows, we explain these different  contributions in more detail, and then evaluate the steady state gain and loss implied by~\eqn{eq:QDBias}.

\begin{figure}
	\begin{center}
		\includegraphics[width=\columnwidth]{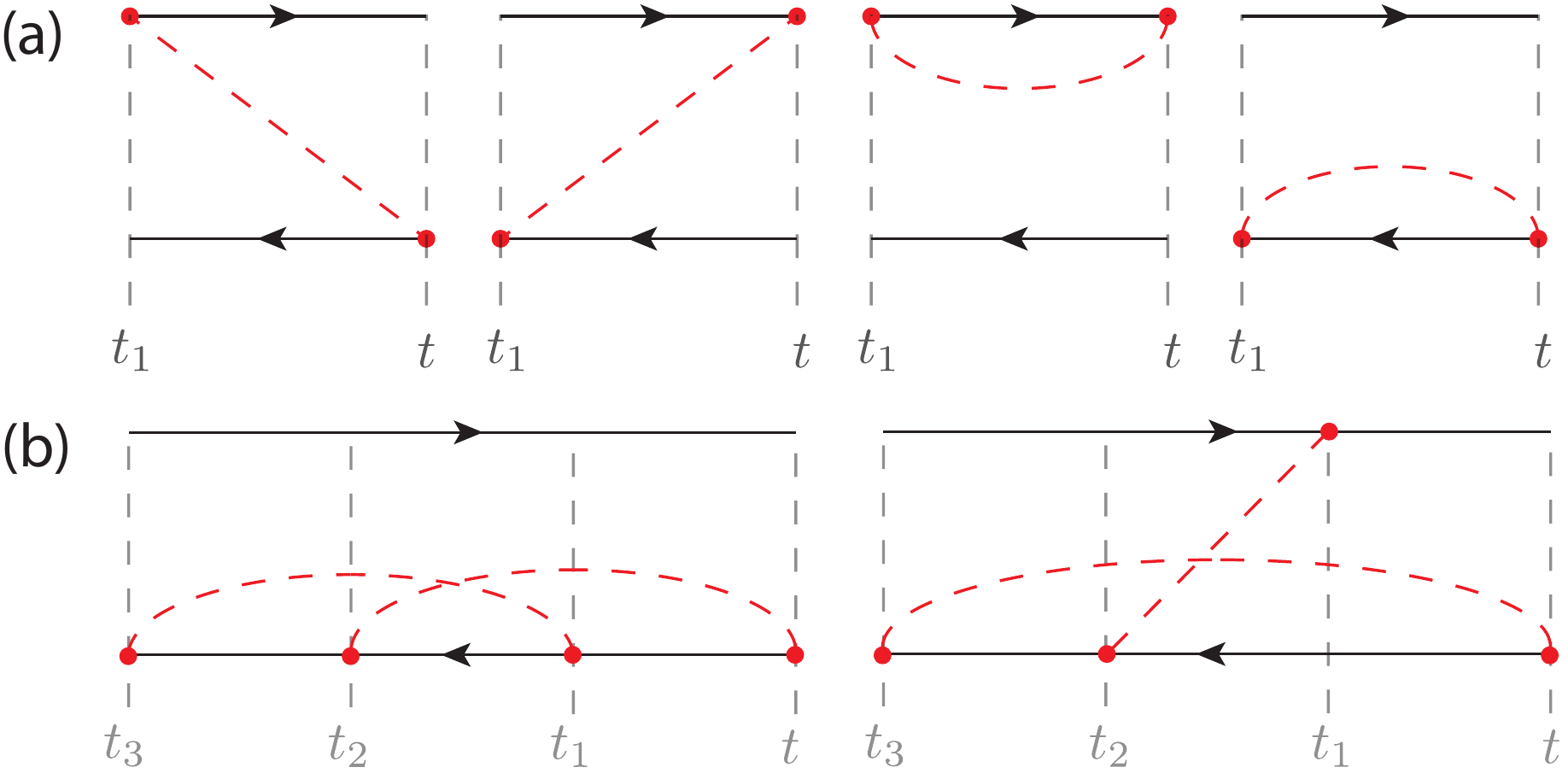}
		\caption{(Color online) (a) Diagrams representing all 2nd order terms in the Keldysh self-energy.  (b) Two examples of irreducible fourth-order Keldysh diagrams. 
			All 32 irreducible diagrams at this order can be generated from the two shown by swapping  sub-sets of interaction vertices between the lower and upper lines.
			For more details, see Ref.~\onlinecite{Mueller:PRA:2016}.
		} 
		\label{fig:keldysh}
	\end{center}
\end{figure}

The second-order dispersive and dissipative terms generated by phonon  and cavity coupling are ~\cite{QuantumNoise, MesoscopicQuantumOptics} 
\begin{eqnarray}
	\mathcal L_{2}\rho&=&-i \comm{H_2}{\rho} + \gamma_{\downarrow,2}\diss{\sigma_-}\rho+\gamma_{\uparrow,2}\diss{\sigma_+}\rho + \gamma_{\varphi,2} \diss{\sigma_z}\rho\nn\\
		&&{}+ \kappa_{-,r}\diss{a}\rho+\kappa_{+,r}\diss{a\hc}\rho,
	\label{eqn:2ndorder}
\end{eqnarray}
where 
\mbox{$\gamma_{\downarrow,2} =  {\sin^{2}\theta}\, C(\omega_{q})/2$}, \mbox{$ \gamma_{\uparrow,2} =  \sin^{2}{\theta} \,C(-\omega_{q})/2$} and
\mbox{$\gamma_{\varphi,2} =  \cos^{2}{\theta}\, C(0) /2$} depend on  the bath spectral function, $C(\omega)$, which is the Fourier transform of the bath correlation function, $\langle X(t)X(0) \rangle$~\cite{MesoscopicQuantumOptics, Mueller:PRA:2016}, 
arising from coupling to the bosonic environment, \mbox{$H_2=\bar \chi \sigma_z(1+2a\hc a)$}, 
\mbox{$\bar\chi = g^2{\sin^2\theta}\, \omega_{q} / (4\omega_{q}^2-4\omega_{r}^2)$} is the dispersive shift between the DQD and resonator,
and $\kappa_{-,r}=\kappa ( n_{\text{th}}+1)$ and $\kappa_{+,r}=\kappa \, n_{\text{th}}$ depend on the cavity decay  rate $\kappa$ and the thermal  population \mbox{$n_{\text{th}}=1/(e^{\beta \omega_r}-1)$}, with $\beta = 1/k_{B} T$.  
Lindblad superoperators are defined as \mbox{$\diss{\mathcal O}\rho=\mathcal O \rho \mathcal O\hc-\{\mathcal O\hc \mathcal O,\rho \}/2$}.  
The first line of~\eqn{eqn:2ndorder}, which includes the dispersive DQD-resonator shift and phonon-induced relaxation, excitation and dephasing, can be obtained by standard quantum optical methods~\cite{Stace:PRL:2005, QuantumOptics}, 
or equivalently, by evaluating  the second-order Keldysh diagrams for the DQD-resonator  and DQD-phonon interaction, shown in~\fig{fig:keldysh}a.

%%%%%%%%%%%%
% Derivation
%%%%%%%%%%%%
%\paragraph{4th order correlated decay terms} - 
The dissipators that arise at higher order are not part of the standard `canon' of Lindblad superoperators in the quantum optical master equation. %Recent theory has identified a few such terms using a polaron transformation \cite{taylor}.  
At a given perturbative order, the complete set can be found by evaluating the irreducible Keldysh self-energy diagrams at that order. The third-order contributions vanish.  There are 32 irreducible diagrams  at fourth order, two examples of which are shown in~\fig{fig:keldysh}b.
In practise, each vertex in a diagram needs to be decomposed into the 9 different Fourier components represented in \eqn{eqn:HI}, so that at $4^\textrm{th}$ order, there are up to $32\times 9^4$ different diagrams to integrate.

These integrals can be evaluated analytically, and grouped into Lindblad dissipators with associated rates.  %we separate each diagram into the fourier decomposition of 
This calculation is described in detail in~\cite{Mueller:PRA:2016}, and results in a total of 21 individual Lindblad terms.  As with $\L_2$, the rates depend on the bath spectral function, $C(\omega)$, and its derivatives~\cite{Shnirman:2016, Mueller:PRA:2016}, evaluated at system frequencies 
$\omega = 0, \pm\omega_{q}, \pm \omega_{r}, \pm (\omega_{q} \pm \omega_{r})$~\cite{Stace:PRL:2005}. 
			
For the purposes of this Letter, we find  six of the 21 Lindblad dissipators represent the dominant contribution  to the correlated DQD-resonator decay, and thus to gain and loss in the resonator field. These are
\begin{subequations}	
\begin{align}
	\mathcal L_{4} \bar \rho = {}& \hphantom{{}+{}}
	\gamma_{\downarrow +}^{(\omega_q-\omega_r)} \diss{\sigma_{-}a\hc}\rho + \gamma_{\downarrow -}^{(\omega_q+\omega_r)} \diss{\sigma_{-}a}\rho \label{eq:4thMEa}\\
		&{}+ \gamma_{\uparrow +}^{(-\omega_q-\omega_r)} \diss{\sigma_{+}a\hc}\rho + \gamma_{\uparrow-}^{(-\omega_q+\omega_r)} \diss{\sigma_{+}a}\rho \label{eq:4thMEb}\\
		&{}+ \gamma_{\varphi +}^{(-\omega_r)} \diss{\sigma_{z} a\hc}\rho + \gamma_{\varphi -}^{(\omega_r)} \diss{\sigma_{z}a}\rho,\label{eq:4thMEc}
\end{align}
\label{eq:4thME}
\end{subequations}
where the rates are given by
\begin{align}
	\gamma_{\downarrow+}^{(\omega_q-\omega_r)} &=  g^{2}\cos^{2}\theta \frac{\omega_q^2\sin^{2}\theta}{2\omega_{r}^{2}(\omega_{q} - \omega_{r})^{2}} C(\omega_{q} - \omega_{r}), \nn\\
	\gamma_{\downarrow-}^{(\omega_q+\omega_r)} &=   g^{2}\cos^{2}\theta \frac{\omega_q^2\sin^{2}\theta}{2\omega_{r}^{2}(\omega_{q} + \omega_{r})^{2}} C(\omega_{q} + \omega_{r}),\nn\\
	\gamma_{\varphi-}^{(\omega_r)} &=  g^{2}\sin^{2}\theta\: 
		\frac{\omega_{q}^{2}\sin^{2}\theta}{2\,(\omega_{q}^2 - \omega_{r}^2)^{2}}
		\: C(\omega_{r}),
	\label{eq:CorrelatedRates}
\end{align}
\mbox{\mbox{$\gamma_{\uparrow -}^{\!(\!-\!\omega_q+\omega_r\!)}\!\!=\! \gamma_{\downarrow +}^{(\omega_q\!-\omega_r\!)} \!e^{-\beta(\omega_{\!q} \!- \omega_{r}\!)}$}, \mbox{$\gamma_{\uparrow +}^{\!(\!-\!\omega_q\!-\omega_r\!)}\!\!=\! \gamma_{\downarrow -}^{(\omega_q+\omega_r\!)}\! e^{-\beta(\omega_{q} + \omega_{r}\!)}$}}
and $\gamma_{\varphi +}^{(-\omega_r)}= \gamma_{\varphi -}^{(\omega_r)} e^{-\beta \omega_{r} }$.
The label $\uparrow(\downarrow)$ denotes a process that excites (relaxes) the DQD, $\varphi$ denotes DQD dephasing, and $+(-)$ denotes photon creation (annihilation).
	
The  terms in~\eqn{eq:4thMEa},b have been derived elsewhere, based on a canonical  transformation to a polaronic frame in which the  qubit  and the resonator are correlated~\cite{Gullans:PRL:2015}.  
They correspond to processes in which the qubit and the resonator both change state, accompanied by exchange of energy with the phonon bath.  This is illustrated in the first two panels of~\fig{fig:dissipation}c.    
 The final two terms in~\eqn{eq:4thME}c are processes that we believe have not been considered in the Lindblad formalism, and correspond to a DQD-mediated exchange of energy between the resonator and the phonon bath that leaves DQD populations unaffected, illustrated in the last panel of~\fig{fig:dissipation}c.

In the experiments that motivate this Letter,  the external leads couple to the open DQD, inducing a charge-discharge transport cycle, pictured in~\fig{fig:dissipation}b.  We extend the DQD  basis to include the empty state $\ket{\O}$, in which the DQD is uncharged. 
As electrons tunnel between the leads and the DQD, it passes transiently through the empty state.    
This process is described by the incoherent Lindblad superoperator~\cite{Lindblad:CMP:1976, QuantumOptics, Stace:PRL:2005}
\begin{equation}
	\L_{\text{leads}}\rho= \Gamma_{L} \diss{\,{\ket L}{\bra \O}\,}\rho + \Gamma_{R} \diss{\,{\ket \O}{\bra R}\,}\rho.
\end{equation}
For simplicity, we will assume $\Gamma_R = \Gamma_{L} = \Gamma$. % and $\op P_{R\rightarrow \O} = \ket \O \bra R$.
Depending on the sign of $\epsilon_q$, the DQD population may become inverted in steady state.

Having established~\eqn{eq:QDBias} describing the dynamics of the correlated DQD-resonator system  we could in principle find the full, correlated steady-state of the system  (in a suitably truncated basis).  
However, we anticipate that the resonator will be close to a coherent state $\ket{\alpha}$, so we proceed by making a mean-field approximation in each of the DQD and resonator subsystems.  This also has the advantage of being computationally  straightforward.  

Thus, we factorise the system density matrix as \mbox{$\rho\approx\rho_r\otimes\rho_q$}, resulting in master equations for each subsystem, mutually coupled through the mean values $\alpha$ and $\mean{\sigma_z}$.  
To proceed, we perform a displacement transformation on the resonator, $a\rightarrow \tilde a+\alpha$~\cite{Gambetta:PRA:2008, Slichter:NJP:2016} and also transform into a rotating frame defined by $ \omega_{d} \tilde a\hc \tilde a$. 
In the displaced resonator frame, and after tracing over the DQD degrees of freedom,~\eqn{eq:QDBias} reduces to a dynamical equation for the resonator 
\begin{align}
	\dot \rho_r = -i \big[\tilde H_{{r}} ,{\rho_{r}}\big] + \kappa_{-} \diss{\tilde a}\rho_{r} + \kappa_{+}\diss{ \tilde a\hc} \rho_{r},\label{eqn:rhor}
\end{align}
where
\begin{align}
\tilde H_{{r}} =& \l(\delta\omega_{r} - 2 \bar\chi \mean{\sigma_{z}}\r) \tilde a\hc \tilde a \nn\\
		&+ \tilde a\hc  \big( \epsilon_{d}/2 + \alpha ( \delta\omega_{r} + 2\bar\chi \mean{\sigma_{z}} -  i \kappa' /2 )	\big)+ \HC,	\label{eq:HDisp}
\end{align}
in which $\delta\omega_{r}=\omega_r-\omega_d$,  $P_{i} = \text{Tr}_{q}\l\{\rho {\ket{i}}{\bra{i}}\r\}$, $\mean{\sigma_{z}}=P_g-P_e$, %  the steady-state populations of the quantum-dot eigenstates and
 and $\kappa'=\kappa'_- -\kappa'_+$ is the DQD-renormalised resonator linewidth, which depends on
 \begin{eqnarray}
	\kappa'_{-} &=& \kappa_{-,r} \!+\! \gamma_{\downarrow -}^{(\omega_q+\omega_r\!)} \!P_{e} \!+\!\gamma_{\uparrow -}^{(-\omega_q+\omega_r\!)} \!P_{g}\! +\! \gamma_{\varphi -}^{(\omega_r)}(1\!-\!P_{\O} ), \label{eq:kappaminus}\\
	\kappa'_{+} &=& \kappa_{+,r} \!+\! \gamma_{\downarrow +}^{(\omega_q-\omega_r\!)} \!P_{e} \!+\!\gamma_{\uparrow +}^{(-\omega_q-\omega_r\!)} \!P_{g} \!+\! \gamma_{\varphi +}^{(-\omega_r)}(1\!-\!P_{\O} ).\nn
\end{eqnarray}
For the parameter regime we consider below, $P_{\O}\ll1$.
	
The coefficient of $ \tilde a\hc$ in~\eqn{eq:HDisp} describes effective driving of the displaced resonator mode $\tilde a$. % and in general lead to a non-zero expectation value of the resonator operators $\mean{\tilde a}$.
We self-consistently choose the displaced frame to eliminate the effective driving, so that
\begin{align}
	 \alpha = - { \epsilon_{d} }/({ 2\,\delta\omega_{r}' - i  \kappa' })\,,\label{Eq:AlphaEigen}
\end{align} where $\delta\omega_r' = \delta\omega_{r} +2\bar\chi \mean{\sigma_{z}}$ is the DQD-renormalised detuning. 
With this choice,~\eqn{eqn:rhor} describes an effective undriven resonator, which will relax to a low energy state with $\langle \tilde a\hc\rangle=\langle\tilde a\rangle=0$ and $\langle \tilde a\hc\tilde a\rangle\ll|\alpha|^2$, as long as $\kappa'>0$. % subject to relaxation  and excitation rates $\kappa'_{\pm}$.

\begin{figure}[!t]
	\begin{center}
		\includegraphics[width=\columnwidth]{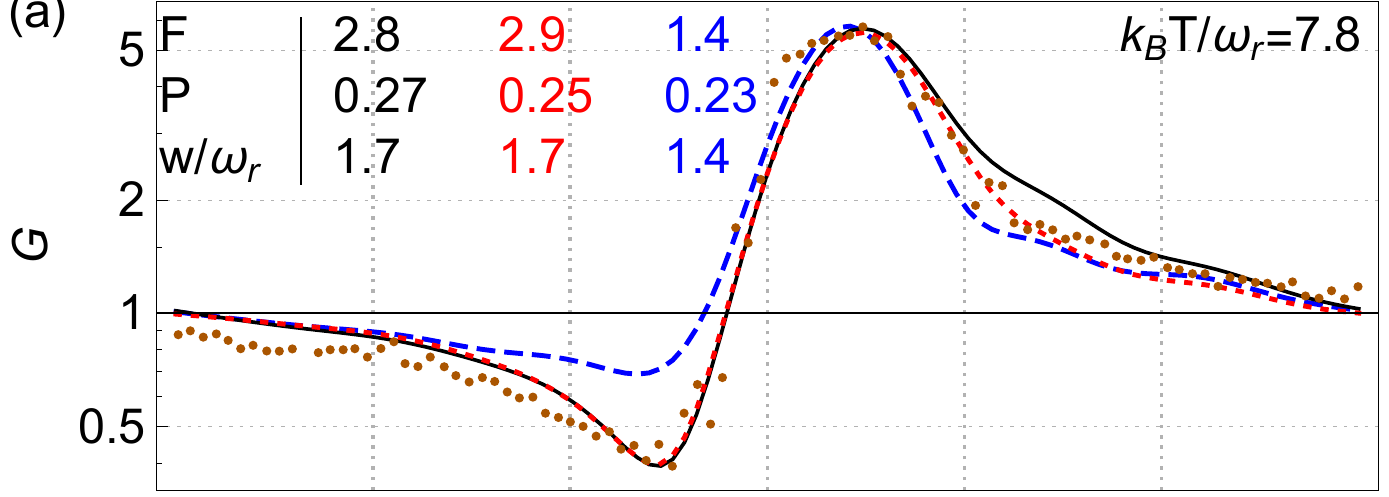}
		\includegraphics[width=\columnwidth]{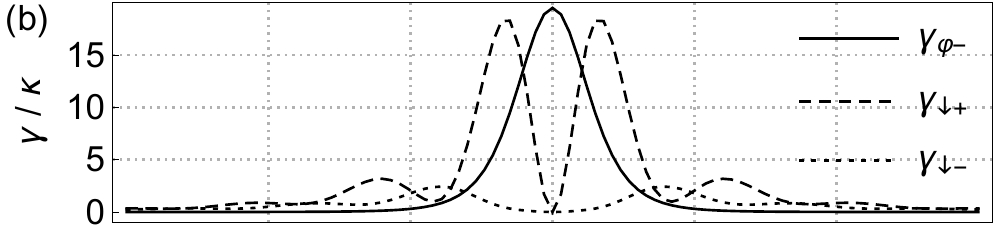}
		\includegraphics[width=\columnwidth]{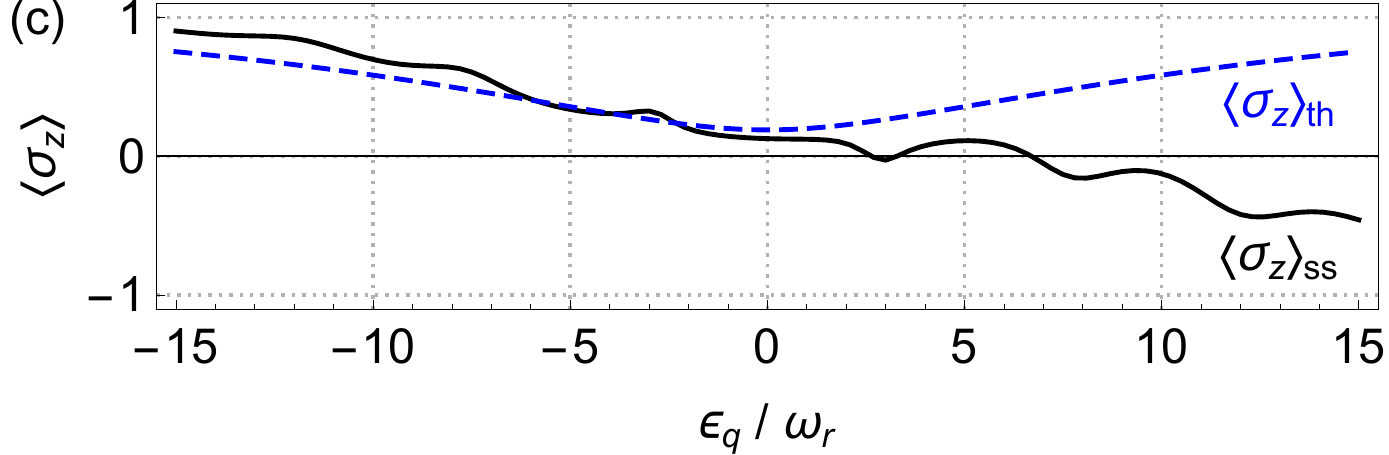}
		\caption{(Color online) {\textbf{(a)}} Logarithmic plot of the microwave power  gain, $G$, versus DQD bias, $\epsilon_{q}$, for  \mbox{$k_{B}T/\omega_{r}=7.8$ (corresponding to $T=3$~K)}.  
			Points are experimental data extracted from Ref.~\onlinecite{Liu:PRL:2014}. 
			The blue-dashed theory curve is generated using terms in the first two lines of~\eqn{eq:4thME}, corresponding to the polaron transformation used in Ref.~\onlinecite{Gullans:PRL:2015}. 
			The solid-black theory curve is generated using the six terms in~\eqn{eq:4thME}. The dotted red curve includes additional terms discussed in~\cite{Mueller:PRA:2016}. 
			\textbf{(b)} Correlated rates, $\gamma_{\downarrow+}$ (dashed), $\gamma_{\downarrow -}$ (dotted), and $\gamma_{\varphi-}$ (solid) from~\eqn{eq:CorrelatedRates} corresponding to the black curve in panel (a). 
			\textbf{(c)} Qubit steady state population $\mean{\sigma_{z}}_{ss}$ (solid, black), compared to thermal population of the DQD, \mbox{$\langle\sigma_z\rangle_\textrm{th}$} (dashed, blue). 
			Common parameters for all panels: $ \omega_{d}/\omega_{r}= 1$, $g/\omega_{r}=0.0125$, $\Delta_{q}/\omega_{r} = 3$, $\kappa/\omega_{r}=52 \times10^{-6}$, $\Gamma/\omega_{r}=0.34$, as in Ref.~\onlinecite{Gullans:PRL:2015}. 
			} 
		\label{fig:gain}
	\end{center}
\end{figure}

%\begin{figure}[!t]
%	\begin{center}
%		\includegraphics[width=\columnwidth]{GainLogPlot}
%		\includegraphics[width=\columnwidth]{rates}
%		\includegraphics[width=\columnwidth]{SigmaZ}
%%		\includegraphics[width=\columnwidth]{GainLogPlot2}
%		\caption{(Color online) (a) Microwave power  gain, $G$, versus DQD bias, $\epsilon_{q}$, for  $k_{B}T/\omega_{r}=7.8$.  
%			Points are experimental data extracted from~\cite{Liu:PRL:2014}. 
%			The blue-dashed theory curve is generated using terms in~\eqn{eq:4thME}a,b, corresponding to the polaron transformation used in~\cite{Gullans:PRL:2015}. 
%			The solid-black theory curve is generated using all terms in~\eqn{eq:4thME}. The dotted red curve includes additional terms discussed in~\cite{Mueller:PRA:2016}.
%			(b) Correlated rates, $\gamma_{\downarrow+}$ (red), $\gamma_{\downarrow -}$ (blue), and $\gamma_{\varphi-}$ (black) from~\eqn{eq:CorrelatedRates} corresponding to the black curve in panel (a). 
%			(c) Qubit steady state population $\mean{\sigma_{z}}$ (solid, black), compared to thermal population of the DQD, \mbox{$\langle\sigma_z\rangle_\textrm{th}$} (dashed, blue). 
%%			 (d) Gain assuming higher temperature, $k_{B}T/\omega_{r}=23.4$. We have adjusted $\mathsf F$ compared to panel (a) to match the gain peak height. 
%%			 Common parameters for all panels: $ \omega_{d}/\omega_{r}= 1$, $g/\omega_{r}=0.0125$, $\Delta_{q}/\omega_{r} = 3$, $\kappa/\omega_{r}=325 \times10^{-6}$, $\Gamma/\omega_{r}=2.125$, as in Ref.~\onlinecite{Gullans:PRL:2015}. 
%			} 
%		\label{fig:gain}
%	\end{center}
%\end{figure}

To  find $P_{e,g}$, we trace~\eqn{eq:QDBias} over resonator degrees of freedom, so that it  reduces to the DQD master equation 
\begin{align}
	\dot \rho_q =& -i \big[{\tilde H_q},{\rho_{q}}\big] + \gamma_{\downarrow}\diss{\sigma_{-}}\rho_{q} + \gamma_{\uparrow} \diss{\sigma_{+}}\rho_{q} \nn\\
		&+ \gamma_{\varphi} \diss{\sigma_{z}} \rho_{q} + \mathcal{L}_{\text{leads}} \rho_{q}
	\label{eq:qDotME}
\end{align}
where
\begin{align}
	\tilde H_q &=- \big( \omega_{q} -2 \bar\chi (1+2\abss{\alpha})\big) \sigma_{z}/2,\\
	\gamma_{\downarrow} &= \gamma_{\downarrow,2} + \abss\alpha \gamma_{\downarrow -}^{(\omega_q+\omega_r)} + (\abss{\alpha} + 1) \gamma_{\downarrow +}^{(\omega_q-\omega_r)} \,,\nn\\
	\gamma_{\uparrow} &= \gamma_{\uparrow,2} + \abss\alpha \gamma_{\uparrow -}^{(-\omega_q+\omega_r)} + (\abss{\alpha} + 1) \gamma_{\uparrow +}^{(-\omega_q-\omega_r)} \,,\nn\\
	\gamma_{\varphi} &= \gamma_{\varphi,2} + \abss{\alpha} (\gamma_{\varphi-}^{(\omega_r)} + \gamma_{\varphi+}^{(-\omega_r)} ) \,.\nn
\end{align}
The steady-state resonator amplitude, $\alpha$, in~\eqn{Eq:AlphaEigen} is  found by solving~\eqn{eq:qDotME} for $P_{e,g}$, subject to $\dot \rho_q=0$. 

The power gain due to the DQD coupling is given by
\begin{align}
	G= \abss{{\alpha}/{\alpha_{0}}}={|2 \,\delta\omega_{r} - i \kappa|^2}/|{2 \,\delta\omega'_{r} - i \kappa'}|^2
	\nn
\end{align}
where $\alpha_{0} = -  { \epsilon_{d} }/({2 \,\delta\omega_{r} - i \kappa})$ is the steady-state resonator field that would be produced in the absence of the DQD coupling, (i.e.\ setting $g=0$).  Below, we set  \mbox{$\delta\omega_{r}=0$}, corresponding to resonant cavity driving.

%%%%%%%%%%%%
% Results
%%%%%%%%%%%%
For comparison with  experimental gain data extracted from \cite{Liu:PRL:2014}, shown as points in \fig{fig:gain}a, we assume the bath spectral function is \mbox{$C(\omega) = J(\omega) \l( n_{\text{th}} + \theta(\omega) \r)$} with \mbox{$J(\omega) = J_{\text{1D}}(\omega) + J_{\text{3D}}(\omega)$}, 
where the  spectral densities for the first phonon mode in the quantum wire and the piezoelectric substrate phonons are given by~\cite{Stace:PRL:2005, Gullans:PRL:2015}
\begin{align}
	\frac{J_{\text{1D}}(\omega)}{\omega_{r}} &= \mathsf{F} \frac{c_{n}}{\omega\, d} \big(1-\cos(\omega \, d /  c_n)\big) \ee^{-\omega^{2} a^{2} / 2 c_{n}^{2}},\nn\\
	\frac{J_{\text{3D}}(\omega)}{\omega_{r}} &= \mathsf{P} \frac{\omega}{\omega_{r}} \big(1-\text{sinc}({\omega\, d/c_{s}})\big) \ee^{-\omega^{2} a^{2}  /2c_{s}^{2}},
%	\frac{J_{1}(\omega)}{\omega_{r}} &= \mathsf{F} \frac{c_{n}}{{\omega }\, d} \big(1-\cos(\omega \, d /  c_n)\big) \ee^{-\omega^{2} a^{2} / 2 c_{n}^{2}},\nn\\
%	\frac{J_{\text{p}}(\omega)}{\omega_{r}} &= \mathsf{P} \frac{\omega}{\omega_{r}} \big(1-\text{sinc}({\omega\, d/c_{s}})\big) \ee^{-\omega^{2} a^{2}  /2c_{s}^{2}}\nn,
\end{align}
with the speed of sound in the quantum wire \mbox{$c_{n}= 4000~m/s$} and in the substrate $c_{s} = 11 000~m/s$, an inter-dot spacing $d = 120$~nm and  nanowire width $a = 25$~nm. 
We treat the dimensionless spectral strengths $\mathsf{F}$ and $\mathsf{P}$ as free parameters, and we tune them so that all theory curves satisfactorily replicate the strong gain peak evident for $\epsilon_q>0$.  The specific values are shown in Fig.~\ref{fig:gain}a. 
As in Ref.~\onlinecite{Gullans:PRL:2015}, we convolve the bare theory gain curves with a normalised gaussian smoothing kernel~$\propto e^{-\epsilon_q^{2}/ 2\mathsf w^{2}}$ to account for low-frequency noise in the gate voltages defining the inter-dot bias~\cite{Petersson:N:2012}.

In addition to experimental data, \fig{fig:gain}a shows three different theoretical curves.  The dashed blue curve includes terms in \eqn{eq:4thME}a,b but not \eqn{eq:4thMEc} (i.e.\ the fourth-order theory restricted to $ \gamma_{\varphi \pm}^{(\mp\omega_r)}=0$), which is equivalent to the polaronic theory in~\cite{Gullans:PRL:2015}.  
The solid black curve further includes terms \eqn{eq:4thMEc}, corresponding to the DQD-mediated photon to phonon interconversion.  
The dotted red curve  includes all 21 fourth-order rates that appear in the derivation of the full master equation, described in our companion paper Ref.~\onlinecite{Mueller:PRA:2016}. 

In \fig{fig:gain}a, there is a clear discrepancy between the polaronic theory (blue, dashed) and the experimental data in the regime \mbox{$\epsilon_q/\omega_r\lesssim0$}: the theory  does not explain the depth of loss (sub-unity gain).  
In contrast, the additional dephasing-mediated processes in \eqn{eq:4thMEc} (black) give rise to enhanced losses beyond the polaronic terms, and are sufficient to quantitatively account  for the entire range of gain and loss observed in the experimental data.  
We have also shown the results of the full fourth-order theory from Ref.~\onlinecite{Mueller:PRA:2016} (red dotted). In the parameter regime described here, the difference between the latter two theories is negligible, apart from a rescaling of $\mathsf{F}$ and $\mathsf{P}$.

\fig{fig:gain}b plots the rates in \eqn{eq:CorrelatedRates}, and shows clearly that the dephasing-assisted  loss rate (black), $\gamma_{\varphi-}^{(\omega_r)}$, is significant compared to the other correlated decay processes, $\gamma_{\downarrow \pm}^{(\omega_q\mp\omega_r)}$ (red and blue).  This is the main reason for the difference between the theory curves in \fig{fig:gain}a.  Since the full fourth-order theory accounts for the experimental data,  we conclude that dephasing-assisted loss is a substantial contribution to the dynamics of the system.

\fig{fig:gain}c shows the steady-state DQD population imbalance, $\langle\sigma_z\rangle$ (black), compared to the thermal equilibrium value, $\langle\sigma_z\rangle_\textrm{th}=\Tr\{\sigma_z e^{-\beta H_S}\}/\mathcal{Z}$ (blue).  In conventional gain/loss models, positive values of the difference  $\langle\sigma_z\rangle_\textrm{th}-\langle\sigma_z\rangle$ (i.e.\ population inversion) drive gain, while negative differences drive loss.  This is manifest in $\kappa'_{\pm}$, where enhancement of $P_e$ leads to an increase of $\kappa'_{+}$, a corresponding decrease in $\kappa'$, and thus an overall increase of $\alpha$.  In contrast, the dephasing-assisted loss contribution to \eqn{eq:kappaminus} is independent of the state of the DQD in the limit that $P_{\O}\ll 1$, which is the case here. 

%\fig{fig:gain}a exhibits residual discrepancies between the experimental data and the theory for $\epsilon_q/\omega_{r}\lesssim-3$.  We note that these can largely be accounted for by increasing the bath temperature to $k_{B}T/\omega_{r}=23.4$, as shown in Fig.~\ref{fig:gain}d. 
%The high local phonon temperature implied by \fig{fig:gain}d ($T= 9$K) may arise from local heating effects due to the current through the DQD.
  
We conclude that the new dephasing-mediated gain and loss Lindblad superoperators in~\eqn{eq:4thMEc} %, which were  derived in our companion paper~\cite{Mueller:PRA:2016}, 
account for substantial additional loss observed in recent experiments.  
These terms were derived in our companion paper using Keldysh diagrammatic techniques~\cite{Mueller:PRA:2016}, and arise at the same order of perturbation theory as other terms previously derived using a  polaron transformation.  
Synthesising Lindblad and Keldysh techniques to derive higher order dissipative terms is thus a powerful approach to a consistent, quantitative understanding of quantum phenomena in mesoscopic  systems.  
Lifting the simplifying mean-field approximation to study the effects of correlations between the DQD and resonator will be the subject of future work.

%%%%%%%%%%%%
% Conclusion
%%%%%%%%%%%%

\begin{acknowledgements}
	We thank J.~H.~Cole, A.~Doherty, M.~Marthaler, J.~Petta, A.~Shnirman, and J.~Taylor for discussions. 
\end{acknowledgements}

\bibliography{../QDotPhonon}

\end{document}